\documentclass[conference]{IEEEtran}
\IEEEoverridecommandlockouts
\usepackage{cite}
\usepackage{amsmath,amssymb,amsfonts}
\usepackage{algorithmic}
\usepackage{graphicx}
\usepackage{textcomp}
\usepackage{xcolor}
\usepackage[T1]{fontenc} 
\def\BibTeX{{\rm B\kern-.05em{\sc i\kern-.025em b}\kern-.08em
    T\kern-.1667em\lower.7ex\hbox{E}\kern-.125emX}}
\usepackage{fancyhdr}

\begin{document}

\title{Technologies and Computing Paradigms: Beyond Moore's law?}

\author{\IEEEauthorblockN{ Daniel Etiemble}
\IEEEauthorblockA{\textit{Computer Science Laboratory (LISN)} \\
\textit{ University of Paris Saclay}\\
Gif-sur-Yvette, France \\
de@lri.fr}

}

\maketitle

\begin{abstract}
As it is pretty sure that Moore's law will end some day, questioning about the post-Moore era is more than interesting. Similarly, looking for new computing paradigms that could provide solutions is important. Revisiting the history of digital electronics since the 60's provide significant insights on the conditions for the success of a new emerging technology to replace the currently dominant one. Specifically, the past shows when constraints and « walls » have contribute to evolution through improved technical techniques and when they have provoked changes of technologies (evolution versus breakthrough). The main criteria for a new technology or a new computing paradigm is a significant performance improvement (at least one order of magnitude). Cost, space requirement, power and scalability are the other important parameters. 

\end{abstract}

\section{Introduction}\label{sec1}

\begin{figure}[htbp]
\centerline{\includegraphics  [width = 8 cm]{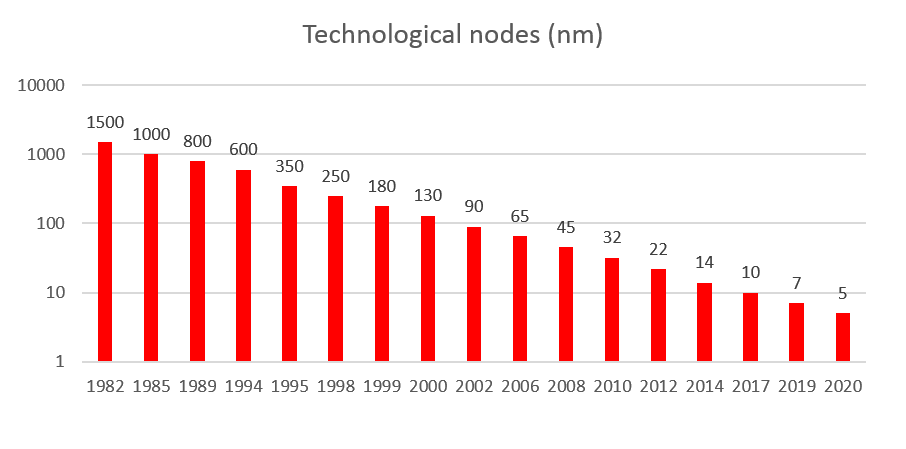}}
\caption{CMOS technological nodes}
\label{figurenode}
\end{figure}

For many dozen of years, the evolution of digital electronics has been driven by Moore’s law that states that the number of transistors in a circuit doubles every N months, even if the process is slowing down, N increasing from 12 to 18 to 24. As a result of Moore’s law, new generations of CMOS/FinFET technologies, called a technological node, have been regularly launched. The different successive nodes are shown in Fig. \ref{figurenode}.

However, everybody agrees on the fact that Moore’s law will end some time, even if it is quite difficult to predict when. 1-nm node is often announced as the last one. So, the big question is : What will be the future of computing beyond Moore’s Law \cite{Shalf}. This question can be decomposed into several ones:
\begin{itemize}
\item Will there be a new technology to continue the exponential performance increase ? 
\item Are there promising technologies that could replace CMOS/FinFET ? 
\item Will there be new computing paradigms to overcome the lack of a replacement of last node CMOS technologies ?
\item Does a Post-CMOS era mean a Post-binary era? \cite{Sandhie}

\end{itemize}

Revisiting the history of digital electronics could provide significant insights on the conditions for the success of a new emerging technology to replace the currently dominant one. Specifically, the past shows when constraints and ``walls" have contribute to evolution through improved technical techniques and when they have provoked changes of technologies.

\section{From transistors to Integrated Circuits with bipolar transistors.}
While the concept of field-effect transistors was proposed in 1926, the first working transistor was invented by Shockley in 1948, one year after  Bardeen and Brattain invented the point-contact transistor. Both were awarded the 1956 Nobel Prize in Physics for their discoveries.
The bipolar transistor has been used in different logic families, either as discrete transistors or integrated circuit gates : RTL then DTL then TTL, ECL, I2L… ECL was the fastest family: it was used in the famous CRAY-1 vector supercomputer.
\subsection{CDC6600 with discrete transistors}
CDC 6600 was a mainframe used from December 1965 to May 1977 that is considered as the first supercomputer. The clock frequency was 10-MHz. This can be compared with the Intel 4004 (first microprocessor) 750 KHz clock frequency launched in 1971. 
The logic circuitry of the CDC6600 is based on resistor-transistor-logic (RTL) in which elementary transistors are used. It should be mentioned that it included some architectural features that became popular later: superscalar execution, multithreading between the central CPU and the peripheral IO processors. Some details can be found in \cite{Eti1}

\subsection{Cray-1 with integrated circuits}
The Cray-1 was launched in 1975 as the first implementation of the vector processor design. The clock frequency was 80 MHz. The Cray-1 used only four different IC types, an ECL dual 5-4 NOR gate (one 5-input, and one 4-input, each with differential output) from Fairchild, another slower MECL 10K 5-4 NOR gate used for address fanout, a 16×4-bit high speed (6 ns) static RAM (SRAM) used for registers and a 1,024×1-bit 48 ns SRAM used for the main memory \cite{Russel}.
\subsection{End of bipolar technology as the dominant technology}
\begin{figure*}[htbp]
\centerline{\includegraphics  [width = 14 cm]{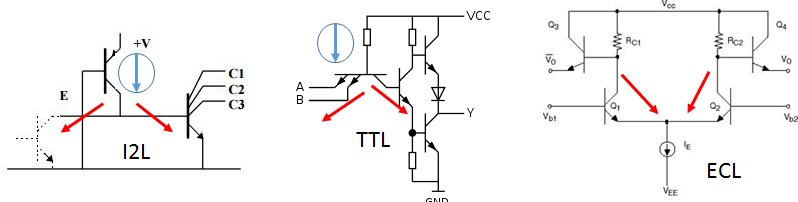}}
\caption{Bipolar circuitries}
\label{fig2}
\end{figure*}

The binary circuitries of I2L, TTL and ECL logic families are presented in Fig. \ref{fig2}. While ECL uses an actual current generator, I2L and TTL use a pseudo current generator. In both cases, the corresponding current flows towards input or output for I2L and TTL, and towards left or right sides of the differential pair for ECL. A significant current contributes to the static power dissipation: it is not needed to consider dynamic power dissipation to understand why bipolar digital circuits have disappeared as soon as MOS technology has  become mature and CMOS circuitry, that had no static power dissipation at that period, was able to replace pMOS and nMOS circuitries.

Power dissipation was the key issue for switching from bipolar to MOS technologies. Bipolar technologies were also far more complicated to integrate than the MOS ones.

Bipolar technology coexisted for a while with MOS technology in a technology called BiCMOS. It was supposed to combine the strengths of the two different process technologies into a single chip: Bipolar transistors offer high speed and gain, which are critical for high-frequency analog sections, whereas CMOS technology excels for constructing simple, low-power logic gates. However, combining two different processes was costly. As soon as MOS technologies were able to provide buffers with sufficient fan-out, the BiCMOS technology was no longer needed for digital circuits.

\section{MOS and CMOS technologies}
\subsection{The first period of MOS technologies}
The MOSFET transistor was invented  by M. M. Atalla and D. Kahng at Bell Labs in 1959, and first presented in 1960. By 1964, MOS chips had reached higher transistor density and lower manufacturing costs than bipolar chips. MOS chips further increased in complexity at a rate predicted by Moore's law, leading to large-scale integration (LSI) with hundreds of transistors on a single MOS chip by the late 1960s.
\subsection{Moore's law}

\begin{figure*}[htbp]
\centerline{\includegraphics  [width = 12 cm]{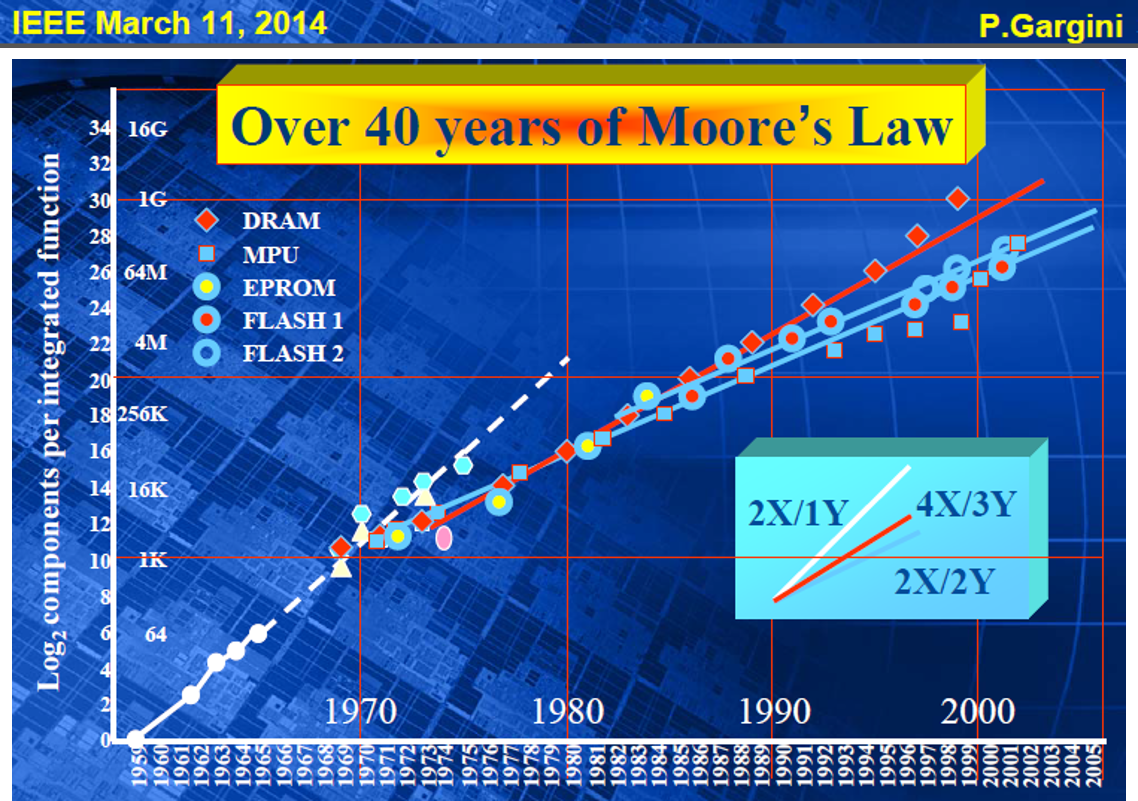}}
\caption{Moore's law presented in \cite{Gardini}}
\label{Moore1}
\end{figure*}

\begin{figure*}[htbp]
\centerline{\includegraphics  [width = 12 cm]{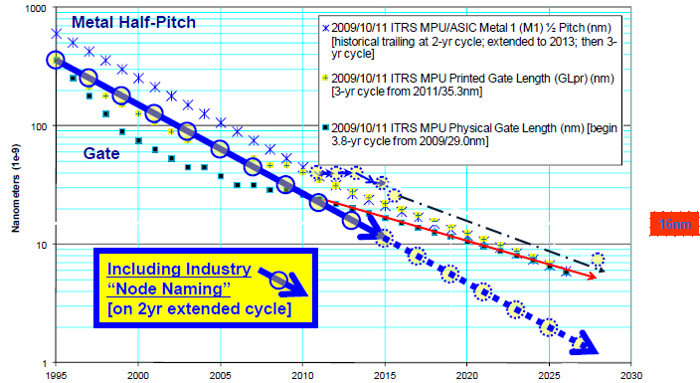}}
\caption{Moore's law slowdown }
\label{Moore2}
\end{figure*}

MOS technologies obey an empirical law, which is a self-fulfilling prophecy, stated in 1965 and known as Moore’s law \cite{Moore}: the number of transistors integrated on a chip doubles every N months. Fig. \ref{Moore1} presents the evolution for DRAM memories, processors (MPU) and three types of read-only memories from 1959 to 2005. The growth rate decreases with years, from a doubling every 12 months to every 18 months (DRAM) and 24 months for microprocessors. The semi-log scale clearly highlights the exponential growth of the number of transistors per chip. Evolution after 2005 is presented in Fig. \ref{Moore2}. 
Even if N tends to increase,  Moore's law is still there in 2022.

\subsection{Scaling rules}
Until mid-2000', MOS technologies follow scaling rules known as Dennard scaling \cite{Dennard}.
With every technology generation:
\begin{itemize}

\item Transistor dimensions could be scaled by –30\% (0.7x).
\item Their area reduces by 50\%, because area is length*width.
\item To keep the electric field constant, the voltage, V, is reduced by 30\% (0.7x), because voltage is field*length.
\item Circuit delays reduce by 30\% (0.7x), because time is length*velocity.
\item The 30\% reduction in delay allows an increase in operating frequency, f, by about 40\% (1.4x), because frequency varies as 1/delay.
\item The 30\% reduction in all distances and related 50\% drop in areas lead to a decrease in capacitance, C, by 30\% (0.7x), because capacitance varies as area /distance.
\item  Power consumption in turn decreases by 50\%, because dynamical power is proportional $V^2f$ (equation \eqref{E1}
\end{itemize}
Therefore, with every new node, area halves and power consumption halves. As the transistor density doubles, power consumption with twice the number of transistors stays the same. 
The limits to this perfect scaling are discussed in \ref{Limits}

\subsection{Two equations}  
We now present two equations that allow to understand the evolution of MOS technologies and the performance of CPUs.
First equation:
\begin{equation}
P_d=P_{s}+P_{d}=V_{dd}*I_{leakage}+\alpha.\Sigma C_i*V_{dd}^{2}*F
\label{E1}
\end{equation}

MOS power dissipation can be broken down into static and dynamic powers. For dynamic power, $V_{dd}$ is the supply voltage, F is the clock frequency, $\Sigma C_i$ is the combined gate and interconnection capacitances while $\alpha$ is the average percentage of circuit capacitances that are switching at each clock cycle: in other words, $\alpha$ is the activity factor of the overall circuit.

Second equation:
\begin{equation}
CPU time = IC*CPI*Tc= \frac{IC}{IPC*F}
\label{E2}
\end{equation}

\begin{itemize}
\item IC is the instruction count of a program
\item CPI is the average Clock cycles Per Instruction and IPC = 1/CPI is the Instruction count Per Clock cycle
\item Tc is the clock cycle time and F=1/Tc is the clock frequency
\end{itemize}

To reduce static power dissipation, a first technique consists in  replacing pMOS and nMOS circuits for which a resistor-like transistor is used as a load by complementary n and p transistors (CMOS). The first MOS circuits used pMOS transistors with 12V power supplies. pMOS circuits were replaced by nMOS circuits that were widely used for computers in the 70's. The first CMOS logic families appeared in the late 60's. At that time, CMOS circuits didn't exhibit any static power dissipation.

\begin{figure}[htbp]
\centerline{\includegraphics  [width = 8 cm]{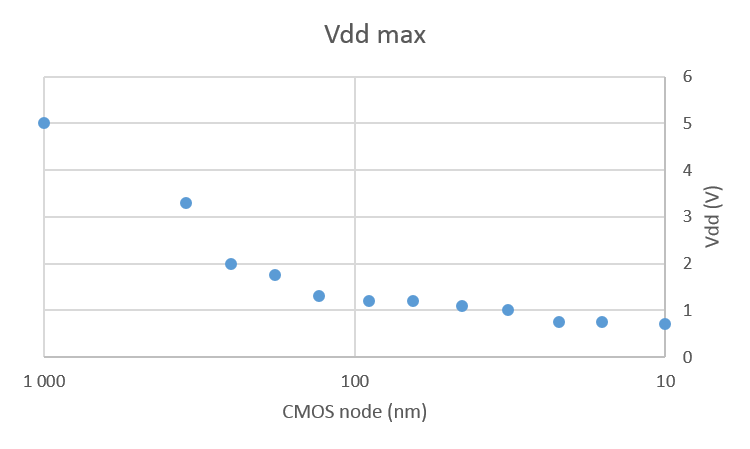}}
\caption{Evolution of power supplies }
\label{Vdd}
\end{figure}

To reduce dynamic power dissipation, one technique consists in reducing $V_{dd}$ as this power dissipation is proportional to $V_{dd}^{2}$. This explains the evolution of power supplies for the successive nodes, as shown in Fig.\ref{Vdd}. 

\subsection{Free lunch}
\subsubsection{Perfect scaling?}
\begin{figure}[htbp]
\centerline{\includegraphics  [width = 8 cm]{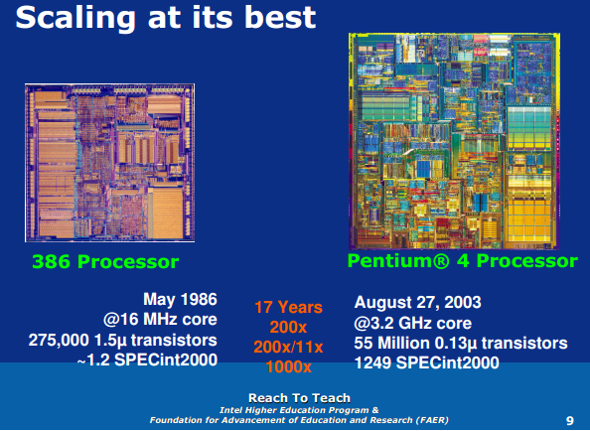}}
\caption{Free lunch according to Intel}
\label{Free}
\end{figure}

The scaling rules of CMOS technology allowed the successive technological nodes (Fig. \ref{figurenode}). These successive nodes leaded to a wonderful period sometimes called ``free lunch". It is illustrated by Fig. \ref{Free}. From 1986 to 2003 (Intel 386 to Intel Pentium 4), there was a 200x increase of clock frequency, 200x increase of transistor count, 11x reduction of transistor channel length and 1000x increase of performance according to Specint2000 benchmark.

\begin{figure}[htbp]
\centerline{\includegraphics  [width = 8 cm]{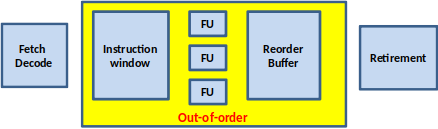}}
\caption{Out-of-order execution in superscalar CPUs} 
\label{ooo}
\end{figure}

\begin{figure}[htbp]
\centerline{\includegraphics  [width = 8 cm]{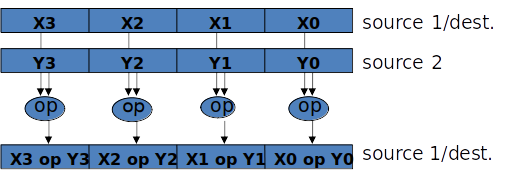}}
\caption{SIMD execution of instructions}
\label{SIMD}
\end{figure}

The performance increase mainly based on three factors:
\begin{itemize}
\item Clock frequency (F): According to the equation \eqref{E2}, execution time is inversely proportional to F. The scaling from one node to the next one reduces the propagation delays and allow to increase clock frequency. 
\item Instructions per Cycle (IPC). According to equation \eqref{E2}, execution time is inversely proportional to IPC (instruction parallelism). This is achieved by replacing  scalar CPUs with a simple pipeline (IPC <1) by superscalar CPUs that launch and execute several instructions per clock, with in-order or out-of order execution. The basic scheme for out-of-order execution is presented in Fig. \ref{ooo} 
\item SIMD parallelism:  SIMD execution of instructions is illustrated in Fig \ref{SIMD}. Arithmetic instructions compute simultaneously several sub-words, reducing the instruction count for programs exhibiting data-parallelism.

\end{itemize}

\subsubsection{Limits of perfect scaling} 
\label{Limits}
\begin{figure*}[htbp]
\centerline{\includegraphics  [width = 12 cm]{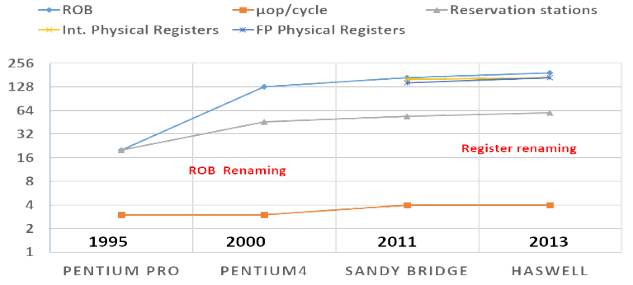}}
\caption{Limits of instruction parallelism}
\label{ILP}
\end{figure*}

It can be noticed that limits of perfect scaling have appeared in this period.
\begin{itemize}
\item Clock frequency: Increasing clock frequency is the simplest way to increase performance. The successive nodes of CMOS technologies lead to a 1.4x decrease in terms of gate delays. This is why there was a yearly 25\% increase in clock frequencies, from 740 kHz (Intel 4004) to 3 GHz (Intel Xeons with 45-nm nodes). But power dissipation is proportional to clock frequency. And increasing clock frequency contributes to the ``Heat Wall" discussed in the next subsection.
\item IPC: there are limits to reachable instruction parallelism, as illustrated in Fig. \ref{ILP}. 
Intel (and AMD) superscalar CPUs break IA-32 and Intel64 CISC instructions down into micro-operations (called µops), which are simpler instructions similar to the RISC instructions of RISC instruction sets. For Intel CPUs, from 1995 to 2013, the hardware resources (physical registers, ROB, reservation stations) have grown significantly. However, this has only resulted in the maximal µop count per cycle to increase from 3 to 4. The additional resources are used to enlarge the number of µops considered for simultaneous launching without changing the maximum number of µops per clock (law of diminushing return).
\end{itemize}

\begin{figure*}[htbp]
\centerline{\includegraphics  [width = 12 cm]{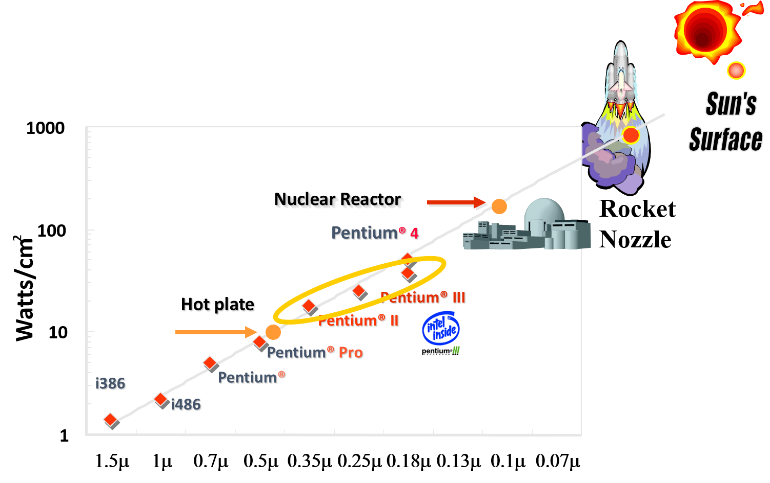}}
\caption{Power density for successive nodes \cite{Pollack}}
\label{HW}
\end{figure*}

\subsection{The Heat Wall}
\subsubsection{Clock Frequency}
Without any change in micro-architectural features, moving from one node to the next allowed a gain in frequency and a subsequent gain in performance even without any CPI change. So, why clock frequencies are limited in the 5-6 GHz range in 2022 while 45-nm nodes already operated at 3-GHz clock frequencies in 2008? With 2022 5-nm technologies, it is quite obvious than tens of GHz would be possible. The answer is simple. Equation \eqref{E1} tells us that the dynamic power dissipation is proportional to the clock frequency and power density exponentially increased from one generation to the next one. This is known as the “heat wall” first coined in \cite{Pollack} and illustrated by Fig. \ref{HW}.  Consequently, the clock frequencies must be limited just to keep the energy budget limits compatible with a safe operation of the components and the typical IC packages and cooling techniques. For higher frequencies, costlier cooling techniques must be used, as shown by the IBM z14 CPU (5.2 GHz) that is water-cooled. The “heat wall” is the main reason for the shift towards multicore architectures, with more, simpler cores replacing single large cores.
\subsubsection{The shift towards multicores}
\begin{figure}[htbp]
\centerline{\includegraphics  [width = 8 cm]{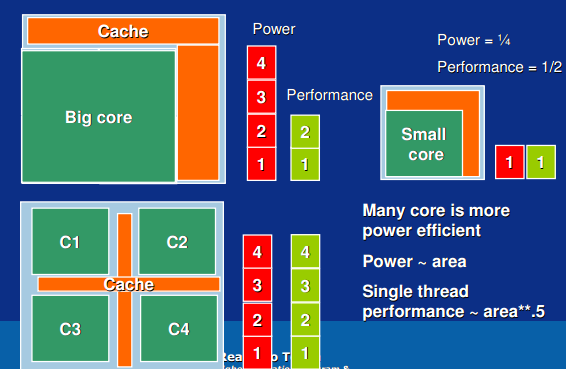}}
\caption{Performance and Power of monocores and multicores}
\label{multicore}
\end{figure}

Fig. \ref{ILP} has outlined the limits of monoprocessors. Fig. \ref{multicore} has been presented by Intel. The figure compares a small core, a large core (4x larger than the small one) and a multicore with four small cores. As a first approximation, power consumption is proportional to the area, while performance is proportional to $\sqrt{area}$. The 4-small core has 2x the performance of the 1-large core with the same power consumption. The 4-small core and the 1-small core have the same performance/power ratio. Obviously, we don't consider here the issues of parallel programming.
 
According to equation \eqref{E2}, parallel processing is also a way to reduce IC. In case of perfect speed-up, IC is divided by the number of cores!

Until mid-2000's, the parallel machines have been confined to the mainframes and the supercomputers. Then, after IBM, Intel and AMD launched the first bi-cores, multicores have become mainstream and are now being used in the whole spectrum of computing devices from the low-end to the high-end application domains.
\subsubsection{Techniques to reduce power dissipation}
Many different techniques have been used to reduce power dissipation, at  technological, circuit or architectural levels.

\begin{figure}[htbp]
\centerline{\includegraphics  [width = 9 cm]{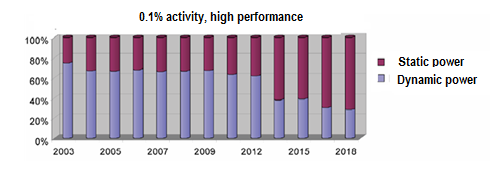}}
\caption{Static versus dynamic power dissipation}
\label{SPD}
\end{figure}

\begin{figure}[htbp]
\centerline{\includegraphics  [width = 6 cm]{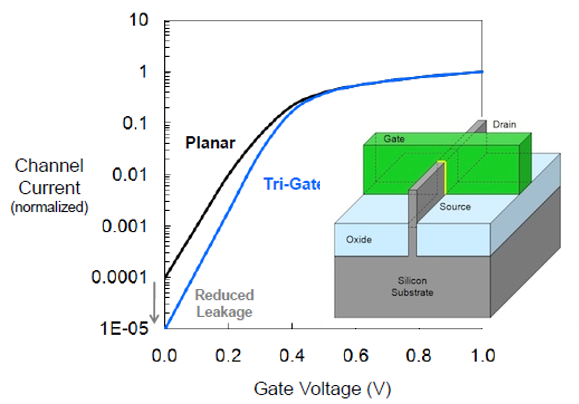}}
\caption{Intel trigate transistors}
\label{TG}
\end{figure}

Old CMOS circuitry had no significant static power consumption. The $V_{dd}$ supply voltage has been reduced for a long time to the minimum value that allows the transistors to operate with correct “on” and “off” states. Until hitting the “heat wall”, CMOS power consumption was mainly due to the clock frequency on one hand, and to $\alpha*\Sigma{C_i}$ that increases with the number of transistors and interconnections that the successive technology nodes allow.
The situation has changed starting with 90-nm nodes due to increased leakage currents. The current going through an “off” transistor is several orders of magnitudes less than the current traversing an “on” transistor. However, the “off” current is not zero. With more and more transistors, the static power consumption becomes increasingly important and can dominate the overall power consumption, as shown in Fig. \ref{SPD}. Intel tri-gate (Fig. \ref{TG}) shows how a technological improvement can reduce the leakage current by one order of magnitude.

At circuit level, a chip can be decomposed into different domains. As power dissipation is proportional to $V_{dd}$, different power domain can be defined to adapt power supplies to the circuit needs. Power dissipation is also proportional to clock frequency. All parts of the circuits do not have the same speed requirements, leading to the decomposition into different clock domains.

There are many other techniques that we do not present here.


\begin{figure*}[htbp]
\centerline{\includegraphics  [width = 14 cm]{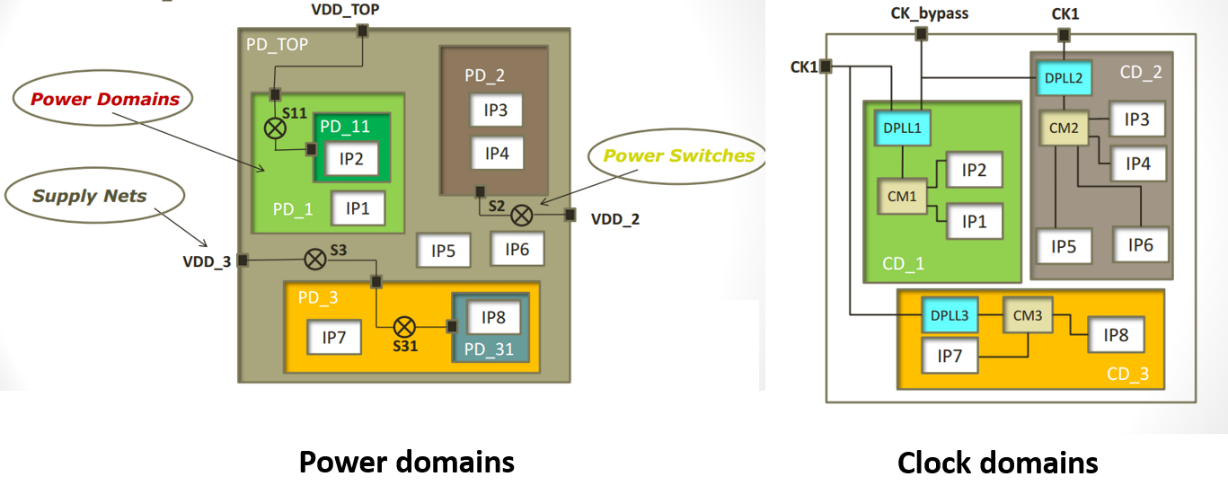}}
\caption{Domain decomposition}
\label{DD}
\end{figure*}

\begin{figure*}[htbp]
\centerline{\includegraphics  [width = 12 cm]{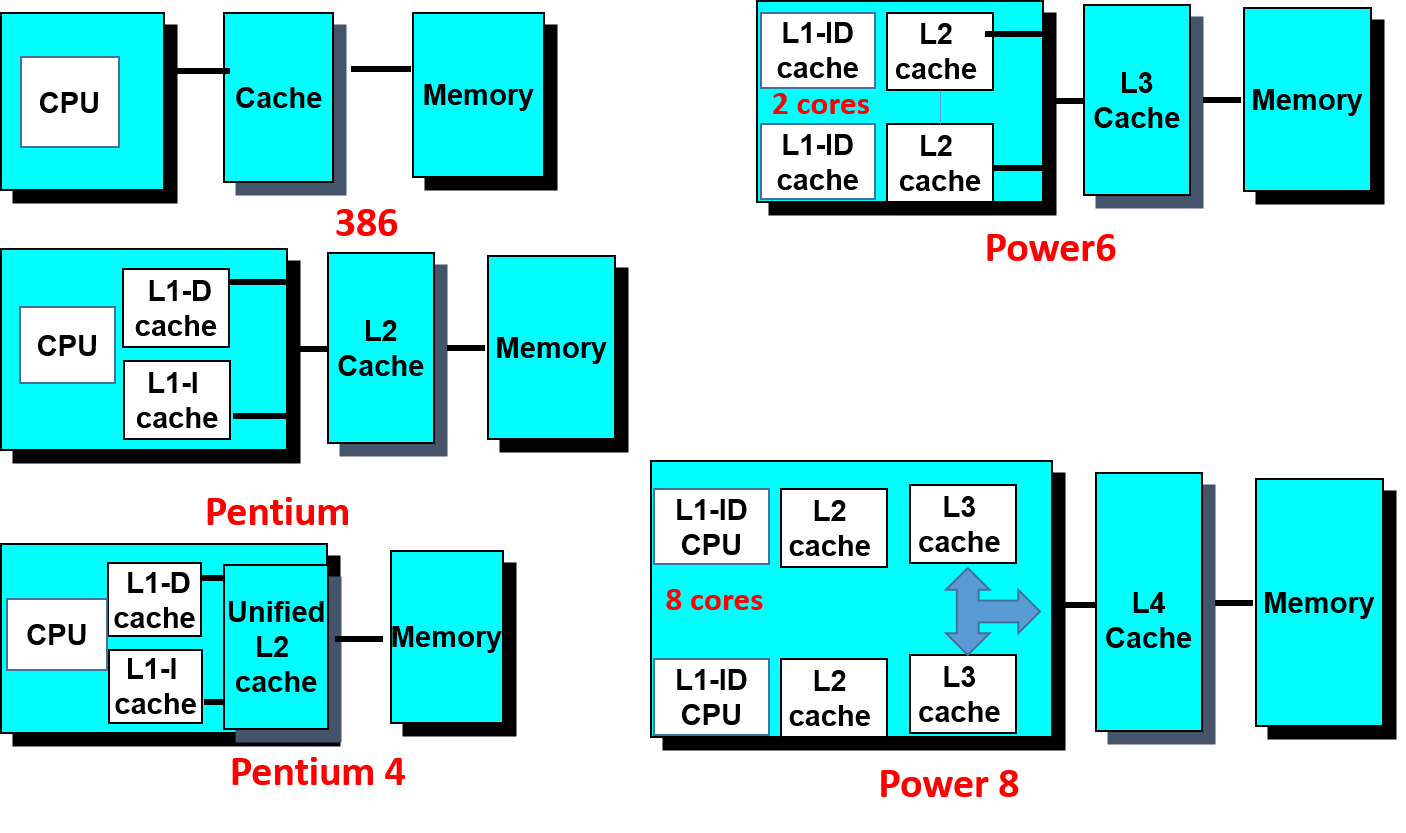}}
\caption{Evolution of cache hierarchies from 1986 to 2014}
\label{MW}
\end{figure*}

\begin{figure*}[htbp]
\centerline{\includegraphics  [width = 14 cm]{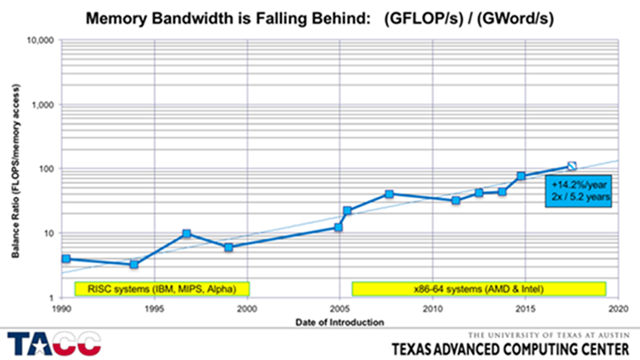}}
\caption{Computations versus Memory Accesses}
\label{Memorybandwith}
\end{figure*}

\subsection{The memory wall}
Processors and DRAM memories both use semiconductor technologies, but they are different. Performance of DRAM access times and bandwidth depend on the DRAM sizes. This situation has leaded to a situation coined as the ``Memory Wall", which is a growing gap between processor and memory performance.
\subsubsection{Caches} 
One popular technique consists in using different levels of cache, with a growing complexity, as illustrated in Fig. \ref{MW} for microprocessors launched between 1986 (Intel 386) and 2014 (IBM Power 8). The number of cache levels increases from 1 to 4, only the last level being off-chip. For multicores with shared memory organization (such as the commercial multicore multiprocessors), cache coherency issues increase significantly. We do not consider here the distributed memory architectures.
\subsubsection{Memory bandwiths}
The second technique consists in improving memory bandwidths. Fig. \ref{Memorybandwith}  presented by Texas Instruments shows the evolution of computing capability (GFLOP/s)/memory bandwith (GWORD/s. More cores in multicores need more memory bandwith, which increases less than the computing capabilities. Successive generations of DDRAMs and HBMs try to improve bandwith. Improving memory bandwith is a key issue.
As an example, the successive generations of IBM Power Systems shows a significant increase of maximal bandwith: 65 GB/s (Power7 in 2010), 210 GB/s (Power8 in 2014), 650 GB/s (Power9 in 2020), 800 GB/s (Power10 in 2021).

\subsubsection{Processing in Memory}
Computing with Memory suppresses the drawbacks of transfers between CPU and Main Memory through a hierarchy of caches. This technique has been studied for a while, and commercial devices are now proposed for applications such as Inference for AI.

\subsection{Discussion on MOS technologies trends}
The history of MOS technologies since the 60's and the corresponding evolution of electronics devices, including microprocessors and computer architecture would need several books. In this paper, we have broadly sketched what we believe to be the salient elements of this history. 

One point is sure. The evolution we described is based on Moore’s law, i.e., an exponential function. As quoted by G. Moore in 2003, “No Exponential is Forever: But “Forever” Can Be Delayed”. The same idea is developed by P. Gardini, Intel Fellow and chairman of ITRS \cite{Gardini}
\begin{itemize}
\item First lesson: “Predictors of Engineering Limits have Always been Proven Wrong by The Right Improvements”.
\item Second lesson: “It Would be Wrong to Believe that the Right Fundamental Limits Don’t Exist”.
\end{itemize}

There had been many examples on how the ``first lesson" has been verified:
\begin{itemize}
\item Reduction of $V_{dd}$
\item Trigate transistors to limit leakage currents
\item Switch from monoprocessors to multicores
\item Etc.
\end{itemize}
``Second lesson" outlines the limits of Moore's law. However, it can be observed than new technical advances are still announced:
\begin{itemize}
\item In May 2021, IBM announced the first 2-nm chip technology \cite{IBM1} that allows to integrate 50 billions transistors.
\item In December 2021, IBM and Samsung announced the VTFET that is a vertical FET when the FinFET transistor is an horizontal one. Paper \cite{IBM2} presents some features of this new transistor: ``VTFET nanosheet and scaled finFET device simulation results are compared at the same footprint and at an aggressive sub-45nm gate pitch. VTFET nanosheets provides 2x performance of the scaled finFET at equivalent power due to VTFET maintaining good electrostatics and parasitics while finFET performance is impacted by severe scaling constraints. Or VTFET could provide as much as 85\% power reduction compared to the scaled finFET architecture as compared at an equivalent frequency on the extrapolated power-performance curves"
\end{itemize}
Whether these announces are evolution or breakthrough is a question. Will they be an actual path to 1-nm node is an open question. However, this shows that research of Right Improvements is continuing.

\begin{figure*}[htbp]
\centerline{\includegraphics  [width = 14 cm]{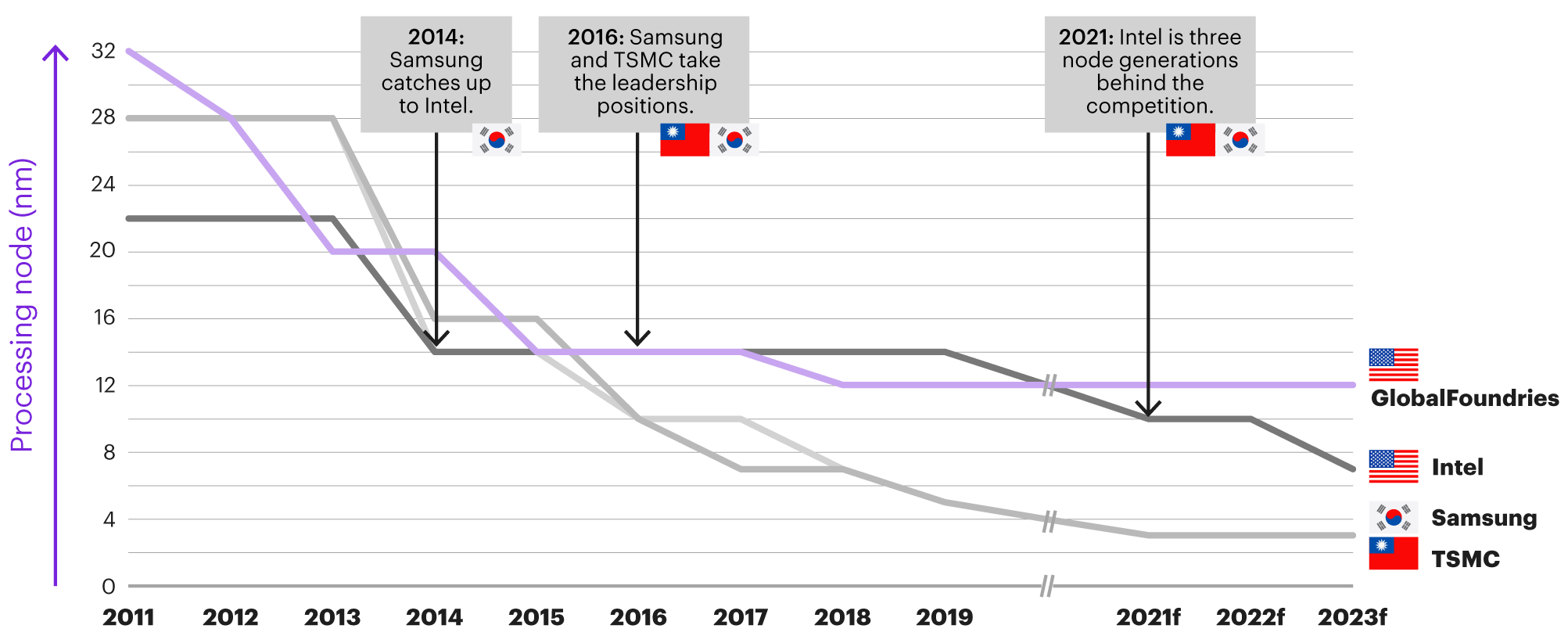}}
\caption{Consequences of Fabs costs}
\label{Fabs}
\end{figure*}

The exponential cost of IC manufacturing, including the costs of fabs, is probably the feature that is the most illustrative of the approaching limits. Better performance means new nodes. Fig. \ref{Fabs} shows the capability of the most significant IC manufacturers to produce new nodes. By 2021, only TMS and Samsung were able to provide the most performing nodes. The Intel troubles with 10-nm nodes also outline the issue. There are technological issues, but also financial issues. In 2021, TMSC dominated 55\% of the market share of semiconductor foundry companies. The second one (Samsung) had 17\% market share. The number of competitors is reducing.

\section{Technological breakthrough: Quantum Computers}
Quantum computers are an actual breakthrough compared to classical computers. There are been a lot of announcements of significant performance gaps:
\begin{itemize}
\item Google announced it has a quantum computer (Sycamore) that is 100 million times faster than any classical computer in its lab.  \cite{Sycamore}
\item In November 2021, IBM announce a quantum computer with 127 qubits \cite{IBM3}
\item Quantum computers have also been developped in China \cite{Choi}
\end{itemize}
Orders of magnitude performance improvement is the key point for the emergence of quantum computers. At the same time, software limits are noticed \cite{Aaronson}. 
\subsection{Algorithms}
Several quantum algorithms would solve specific problems, such as factoring integers, exponentially faster than any known classical algorithm. However, there are many others, such as playing chess, proving theorems, scheduling air flights, for which quantum computers would suffer from the same algorithmic limitations as classical computers. They would surpass conventional computers only slightly \cite{Aaronson}. This is why they would be used as coprocessors for the exponentially faster applications.
\subsection{Hardware}
\begin{figure}[htbp]
\centerline{\includegraphics  [width =4 cm]{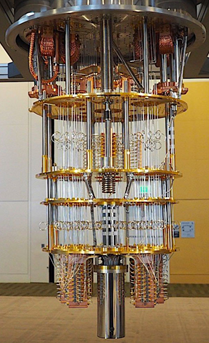}}
\caption{The IBM 50-qubit computer}
\label{QC}
\end{figure}

 Quantum physics is totally different from classical physics, but operational conditions for quantum devices are also completely different from the conventional integrated circuits. We only list a first features of the 50-qubit IBM quantum computer:
\begin{itemize}
\item The qubits are processed at 15 mK. Different stages operate at 4 K, 800 mK, 100 mK and 15 mK, very close to the absolute zero.
\item The quantum processor is located inside a shield to protect it from electromagnetic radiation
\item The coaxial line between the first and second amplifying stages are made out of superconductors.
\item Quantum amplifiers inside  a magnetic shield capture and amplify processor readout signal while minimizing noise.
\end{itemize}
Fig. \ref{QC} shows that this quantum computer has few similitudes with the computing devices that we currently use.  IBM and D-Wave quantum computers operate at temperature close to 0\textsuperscript{o} K. Another approach developed by Paskal operates at room temperature by using neural atoms activated by lasers. 
Quantum computing is a must and the evolution is quite fast. However, even with more and more applications, it will remain a niche. Either with close to 0\textsuperscript{o} K temperatures or with lasers, quantum computers or devices are truly different of the whole world of personal and IoT devices.

\section{Other technologies?}
We now examine some technologies that are not actual potential candidates for Post-Moore era. Some of them have features that make them useful among currently used technologies. However, they have not the needed features in terms of performance gap, integration density, scaling properties, costs, etc. to be considered as candidate for Post Moore era.
\subsection{Magnetic technologies}
Magnetic technology has been extensively used at the beginning of computing history. From 1955 to 1975, magnetic-core memories were the main memory of mainframes. These core memories were non-volatile. They were replaced by the volatile semi-conductor DRAMs. Since the 90's, different types of non volatile magnetic RAMS have been developed and fabricated: Ferroelectric RAM (FRAM), Magnetoresistive RAM (MRAM), Phase-Change memory (PCM)... Some techniques are mature. A 1Gb MRAM has been launched by Everspin. Samsung announces MRAM-based in-memory computing. These developments are very promising. Mature non-volatile memories are perfect candidates to replace SRAM-based caches. In-memory computing is still in the research phase. While these magnetic devices find their place in the current electronics world, they do not seem to be a global alternative to the main stream technologies.

\subsection{Optical computing}
Optical computing has been studied for dozens of year. Paper \cite{Sawchuk} has got 217 paper citations. The conclusion of the abstract was: ``Finally, the current limitations and future needs of optical logic devices and digital optical computing systems are outlined". Optical fibers and devices are widely used for telecommunications. For example, they are the backbone of the Internet infrastructure. But there has not been significant advances for computing. Similar prospects and challenges of optical computing are described in the blog \cite{Optic}. It doesn't mean that breakthroughs are impossible. In \cite{Choi2}, a new optical switch up to 1000x faster than transistors is presented. However, the author Pavlos Lagoudakis, a physicist at the Skolkovo Institute of Science and Technology in Moscow precise: ``It took 40 years for the first electronic transistor to enter a personal computer and the investment of many governments and companies and thousands of researchers and engineers." He says: ``It is often misunderstood how long before a discovery in fundamental physics research takes to enter the market." In other words, 
these advances are far from being a solution for Moore end.

\subsection{CNTFET technologies}
A lot of papers using 32-nm CNTFET technology have been published in the last decade, especially for m-valued circuits. It turns out that this technology lacks of integration density.

In 2012, IBM announced a breakthrough in nanotube computer chip fabrication \cite{Park}. The circuit had 10,000 CNTFETs. The IBM researchers announced: ``Carbon nanotubes have the potential in the development of high-speed and power-efficient logic applications. However, for such technologies to be viable, a high density of semiconducting nanotubes must be placed at precise locations on a substrate'' and ``This new placement technique is readily implemented, involving common chemicals and processes, and provides a platform for future CNTFET experimental studies". Other breakthroughs were announced in 2017 \cite{ibmcntfet}, but it looks like IBM no longer focuses on CNTFET technology.
In 2013, the first carbon nanotube computer has been announced \cite{Shulaker} by a Stanford group. It was a significant advance for this technology. However, this 178 CNTFETs ``one-instruction-set computer" only operated at 1 KHz. The first commercial microprocessor (Intel 4004) had 2300 transistors and ran at 780 KHz in 1971.
In 2019, a 16-bit RISC microprocessor has been built with 14,000 CNFET transistors \cite{Hill} by the same Stanford group. While this is a significant advance for CNTFET technology, we may observe that the Intel 8086 CPU, which was also a 16-bit microprocessor, has been launched in 1978 with 29,000 transistors, more than 40 years ago! In 2019, the largest  transistor count in a commercially available microprocessor was 39.54 billion MOSFETs, in AMD's Zen 2 fabricated using TSMC's 7 nm FinFET semiconductor manufacturing process. In 2020, the largest transistor count in a GPU (NVidia Ampere) was 54 billion transistors with the same 7 nm process. The CNTFET 16-bit microprocessor manufacturing process had 5 metal layers, while the number of metal layers in nano-CMOS technologies ranges from 8 to 15, with a trade-off between integration and cost. 
It does not seem that CNTFET technology can compete with the most recent FinFET technologies.

\section{New computing paradigms}
Within the exponential increase of Integrated Circuits performance, new computing paradigms have also appeared. The main feature is a significant performance improvement.
\subsection{Graphic Processing Units (GPUs)}
\begin{figure*}[htbp]
\centerline{\includegraphics  [width=14 cm]{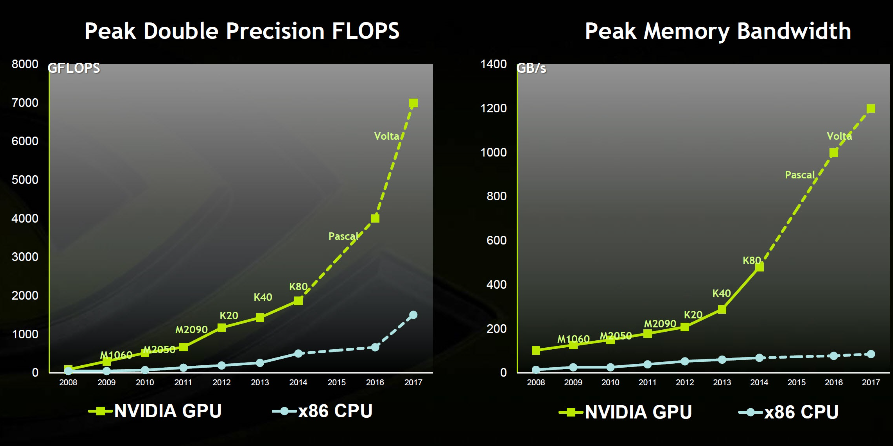}}
\caption{GPU-CPU performance}
\label{GPU}
\end{figure*}
GPUs have initially been designed for computer graphics. In the early 2000's, they started being used for scientific computation. The main reason is shown in Fig. \ref{GPU} presented by NVidia. An order of magnitude difference can be observed both for Peak double precision FLOPS (floating point computation) and Peak Memory Bandwidth. The cohabitation is not simple. CPUs and GPUs have two different computing models. CPU uses large-grain parallelism while GPU uses fine-grain parallelisme to exploit data parallelism. The programming models are different and GPUs use specific application programming interfaces such as CUDA (NVidia) or OpenCL. A GPU operates as a coprocessor. Tranfers between CPU and GPU must be handled. While programming is more complex, the performance advantage is such that 7 out of the top10 positions of the November 2021 TOP500  ranking were supercomputers using GPUs as accelerators.
\subsection{FPGAs}
Field Programmable Gate Arrays (FPGAs) appeared mid 80's as a way to design circuits by programming the interconnects of gates assembled in an array. Thanks to Moore's law, the transistor count of most efficient FPGAs is now tens of billions. Some hardware features have been added to the array of logic cells such as memory blocks, multipliers, hard microprocessors, I/O peripheral circuits... and even AI tensor blocks. They constitute a cost effective approach to get high performance devices without using the ASIC approach. They are often part of Systems on Silicon (SoC). However, FPGAs belongs to the Moore era: their future will depend on Moore's law future.
\subsection{Deep Neural Networks (DNNs)}
\begin{figure*}[htbp]
\centerline{\includegraphics  [width=14 cm]{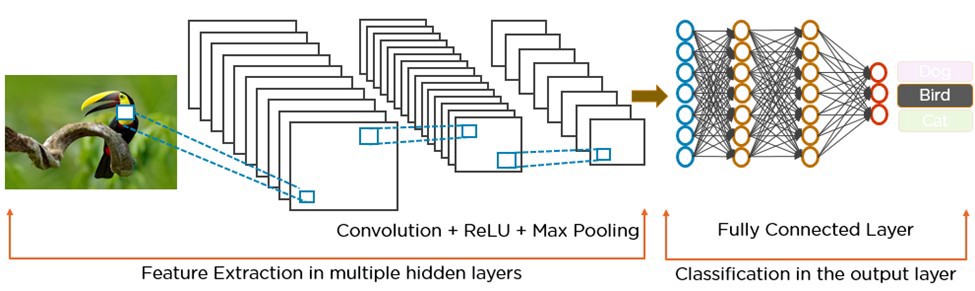}}
\caption{A DNN for image recognition}
\label{DNN}
\end{figure*}

Neural networks were first proposed in 1944 by W. McCullough and W. Pitts, two University of Chicago researchers. Since that period, neural networks have been continuously studied, but the breakthrough arrived in the early 2010's with Deep Neural Networks (DNNs). Fig. \ref{DNN} shows a specific type of DNN called CNN for image recognition. A first step consists in several levels of operators (Convolution, ReLU, Pooling) to reduce the input data set. The second step consists in several levels of fully connected layers to realize the classification. 

There are many types of applications using DNNs such as speech recognition, image recognition, natural langage processing, etc. DNN is a very hot topics in hardware design with:
\begin{itemize}
\item New specific instructions in General Purpose Instruction Set such as Intel64
\item Specialized operators such as Tensor and Tensor cores in NVidia GPUs
 or Intel FPGAs
\item Neuronal processors such as ARM Ethos, Google TPU, Intel Nirvana or Xilinx Versal AI core
\end{itemize}
DNN networks are programmable (parameters of convolutions, pooling ; weights of the neurons). However, this programming is dedicated to each application and is different from programming general purpose microprocessors. DNN are application-specific architectures. 

While DNNs are a very successful approach for a lot of applications, their success is based on Moore's law, i.e. they profit from the huge number of available transistors to improve performance.   

\subsection{Multivalued circuits}
Multivalued circuits are sometimes presented as the solution for the post-CMOS era \cite{Sandhie}. Opposed to GPUs, FPGAs or DNNs, multivalued circuits are not a solution. Assuming a perfect scaling between binary and multivalued circuits performance, the gain would be $log(M)/log(2)$, i.e. x1.585 for ternary circuits or x2 for binary circuits. However, comparing binary and multivalued combinational circuits shows that the multivalued circuits are always more complicated: more interconnects, more chip area, more power, etc \cite{Eti2}. There is a fundamental reason: whatever is the electrical support of the different values (voltage, current, charges) that is used, the different values are totally ordered. At the mathematical level, it means that the Boolean algebra (binary circuits) must be replaced by one variant of the Post algebras.

We illustrate the issue with a small example. Just consider a ternary system in which 0<1<2. All ternary designers use the binary functions NI and PI presented in Table \ref{T2}. It means that the ternary variable is decoded into two binary variables. After some binary computations, the binary result is encoded as a ternary result. These ternary to binary decoding and binary to ternary encoding means that the ternary combinational circuits are always more complex than the binary ones that processes the same amount of information.

Any technology using a totally ordered set of values is faced to this issue. There exists one successful multivalued technology: quantum devices and computers because qubits have multiple unordered values. 

Multivalued circuits are reduced to a very small niche \cite{b2}. In 2022, the only significant use is for Flash Memories: as access time is not critical for these devices used in USB memories, larger access times are compensated by smaller chip areas.
\begin{table}
\centering
\caption{NI and PI binary functions}
\begin{tabular}{|c|c|c|c|c|c|c|c|c|}
  \hline
 &NI&PI\\
  \hline
0&2&2\\
1&0&2\\
2&0&0\\
  \hline
\end{tabular}
\label {T2}
\end{table}

\section{Concluding remarks}
While the end of Moore's law is coming, we examined the lessons of the last 60 years to try to find a successor to the actual dominant semiconductor technology. More precisely, we consider the required conditions for a technology to replace the currently used one. We also consider the conditions for a new computing paradigm to emerge, as a dominant one or as a complement to the usual paradigms.

A first observation that a new technology should bring a significant performance improvement, at least one or several orders of magnitude. Quantum devices and computers are the perfect example. While they operate close to 0°K, while they look more like the ENIAC than to our desktops, laptops or smartphones, they can solve specific problems can would need weeks or years of computation with the best supercomputers. At the same time, they are efficient for specific problems.

Without that huge performance gap, a new technology can be successful if it has  significant features such as cost, size, power and scaling properties. The shift from bipolar technology to MOS technology corresponds to that situation. The scaling rules that we described allows MOS technology, which was slower than bipolar technology at the beginning to finally overcome bipolar technology.

Within the Moore's law era, the same remarks can be made for new computing paradigms. GPUs performance advantage made them good candidates as coprocessors of CPUs. They cannot replace CPUs. The emergence of Deep Neural Networks is another good example. Based on a totally different computing paradigm, they provide exceptional performance on a large range of specific applications, without being a general purpose approach.

Performance, cost, size, power, scalability are the key features needed for a Post Moore technology. Will such a technology be available at the end of the Moore's law ? It is difficult to predict the future! However, it seems even more difficult now to find a good candidate.

\end{document}